\documentclass[showpacs,twocolumn,aps,prl,superscriptaddress]{revtex4}
\usepackage{amsmath}
\usepackage{amsfonts}
\usepackage{amssymb}
\usepackage{amsthm}
\usepackage{graphicx}

\begin{document}
\title{Euler equation of the optimal trajectory for the fastest 
magnetization reversal of nano-magnetic structures}
\author{X. R. Wang}
\affiliation{Physics Department, The Hong Kong University of
Science and Technology, Clear Water Bay, Hong Kong SAR, China}
\author{P. Yan}
\affiliation{Physics Department, The Hong Kong University of
Science and Technology, Clear Water Bay, Hong Kong SAR, China}
\author{J. Lu}
\affiliation{Physics Department, The Hong Kong University of
Science and Technology, Clear Water Bay, Hong Kong SAR, China}
\author{C. He}
\affiliation{Physics Department, The Hong Kong University of
Science and Technology, Clear Water Bay, Hong Kong SAR, China}
\date{\today}

\begin{abstract}
Based on the modified Landau-Lifshitz-Gilbert equation for 
an arbitrary Stoner particle under an external magnetic field 
and a spin-polarized electric current, differential equations 
for the optimal reversal trajectory, along which the 
magnetization reversal is the fastest one among all possible 
reversal routes, are obtained. 
We show that this is a Euler-Lagrange problem with constrains. 
The Euler equation of the optimal trajectory is useful in 
designing a magnetic field pulse and/or a polarized electric 
current pulse in magnetization reversal for two reasons. 
1) It is straightforward to obtain the solution of the Euler 
equation, at least numerically, for a given magnetic 
nano-structure characterized by its magnetic anisotropy energy.
2) After obtaining the optimal reversal trajectory for a given 
magnetic nano-structure, finding a proper field/current pulse 
is an algebraic problem instead of the original nonlinear 
differential equation. 
\end{abstract}
\maketitle
The advent of miniaturization and fabrication of magnetic 
particles with single magnetic domains\cite{Qikun}, called 
Stoner particles, makes the Stoner-Wohlfarth (SW) problem\cite
{book} very relevant to nano-technologies and nano-sciences. 
One current topic in nanomagnetism is the controlled 
manipulation of magnetization dynamics of Stoner particles. 
Magnetization state can be manipulated by a magnetic field\cite
{field,xrw,xrw1} through the usual magnetization-field 
interaction, or by a spin-polarized electric current\cite
{Slon,theory,three,exp,zhang,sun,xrw2} through the so-called 
spin-transfer torque (STT), or by a laser light\cite{Bigot}. 
In terms of applications, controlled manipulation of a 
magnetization by a magnetic field and/or a spin-polarized 
electric current is and would be dominating techniques in 
information storage industry. The magnetization dynamics 
of a Stoner particle under the influence of a magnetic 
field and/or a spin-polarized current is described by 
the so-called nonlinear Landau-Lifshitz-Gilbert (LLG) 
equation that does not have analytical solutions. 
Most of the results obtained so far on magnetization 
reversal are either from a numerical solution\cite{field} 
of a given field/current pulse or based on a specific 
feature\cite{sun,xrw,zhang} of the LLG equation.
Important issues are to lower critical fields/currents 
required to reverse a magnetization\cite{sun} and to design 
a field/current pulse such that the magnetization can be 
switched from one state to another as quickly as 
possible\cite{xrw1,xrw2}. 

Many reversal schemes\cite{sun,xrw,zhang} have been proposed 
and examined. To evaluate different reversal schemes, it is 
important to know the theoretical limits of the critical 
switching field/current, and the optimal field/current pulse 
for the fastest magnetization reversal. The answers to the 
above questions were only known recently for the simplest 
Stoner particles of uniaxial magnetic anisotropy when either 
a magnetic field or a spin-polarized electric current\cite
{xrw1,xrw2} is used to reverse the magnetization. 
Unfortunately, those ideas and approaches, based on the 
rotational symmetry of the system around its easy-axes, 
are not applicable to a non-uniaxial Stoner particle. 
In fact, the method does not even work on an uniaxial Stoner 
particle reversed by a spin-polarized electric current and a 
static magnetic field non-collinear to its easy-axis, a 
possible useful model in current-controlled MRAM (magnetic 
random access memory) devices. 
The solutions (answers) of these questions for an arbitrary 
Stoner particle under combined influences of a magnetic field 
and an electric current are the themes of current work. 
In this letter, the Euler equation of the optimal reversal 
trajectory is derived. For the special case of reversal of 
uniaxial Stoner particles by either a magnetic field or a 
Slonczewski-type of STT, analytical solution of the equation 
can be obtained, which accord with the early results in 
References 5 and 12. In general, the equation can be 
numerically solved easily for a given Stoner particle of a 
known magnetic anisotropy energy. Given a reversal trajectory, 
finding the required field/current pulse is an algebraic 
problem. The generality of the current theory is demonstrated 
on a biaxial Stoner particle.

Consider a Stoner particle of a magnetization $\vec{M}=M
\vec m$ under the influence of an external magnetic field 
$\vec H$ as well as a polarized electric current of $I$, 
where $M$ is the saturated magnetization of the particle 
and $\vec m$ is the unit vector of $\vec M$. Theoretical 
studies\cite{Slon,theory,three} show that the STT $\Gamma$ 
is proportional to the current with following form
\begin{equation}
\Gamma \equiv [\frac{d(\vec{M}V)}{dt}]_{STT} =\frac{\gamma\hbar
I}{\mu_0 e } g(P, \vec{m}\cdot \hat{s}) \vec{m}\times (\vec{m}
\times \hat{s}), \nonumber
\end{equation}
where $\hat{s}$ is the polarization direction of the current. 
In the expression, $V$, $\hbar$, $\mu_0$($=4\pi\times 
10^{-7}N/A^2$), and $e$ denote the volume of the magnetic 
nano-structure, the Planck constant, the vacuum magnetic 
permeability, and the electron charge, respectively. 
$\gamma=2.21\times 10 ^5 (rad/s)/(A/m)$ is the gyromagnetic 
ratio. The exact microscopic formulation of STT is still 
a debating subject\cite{theory,three}. Most of existing 
theories differ themselves in different $g$-functions 
that depend on the degree of the polarization $P$ of the 
current and relative angle between $\vec{m}$ and $\hat{s}$. 
Many experimental investigations\cite{exp} so far 
are consistent with the result of Slonczewski\cite{Slon}, 
$g= 4P^{3/2}/[(1+P)^3(3+\vec{m}\cdot\hat{s})-16P^{3/2}]$, 
which will be assumed in this study whenever we need an 
explicit expression of $g$.

The dynamics of the magnetization $\vec M$ is governed by 
the generalized LLG equation\cite{xrw2},
\begin{equation}\label{LLG}
\frac{d\vec{M}}{dt}=-\gamma\vec{M} \times \vec{H}_{t}+\alpha
\vec{m} \times \frac{d\vec{M}}{dt} +\gamma a_I\vec{M}\times
(\vec{M} \times \hat{s}),
\end{equation}
where $\alpha$ is the phenomenological dimensionless damping 
constant and $a_I=\hbar Ig/(\mu_0 e M^2V)$ is a dimensionless 
parameter of Slonczewski STT\cite{Slon}. $\vec{H}_t$ is the 
usual total effective magnetic field. Notice that $M$ is a 
constant according to Eq. \eqref{LLG}, it is convenient to 
re-write the LLG equation in the dimensionless form, 
\begin{equation}\label{DMLLG}
(1+\alpha^2)\frac{d\vec{m}}{dt}=-\vec{m}\times\vec{h}_1-
\vec{m}\times(\vec{m}\times\vec{h}_2),
\end{equation}
where
$\vec{h}_1 = \vec{h}_t+\alpha a_I\hat{s},$ and 
$\vec{h}_2 = \alpha\vec{h}_t- a_I\hat{s}$. $t$ in 
Eq. \eqref{DMLLG} is in the units of $(\gamma M)^{-1}$. 
Both magnetization and magnetic field are in the units of $M$. 
The total field $\vec{h}_t\equiv \vec{H}_t/M=\vec{h}+
\vec{h}_i$ includes both the applied magnetic field $\vec{h}=
\vec{H}/M$ and the internal field $\vec{h}_i$ due to the 
magnetic anisotropic energy density 
$w(\vec{m})$, $\vec{h}_i=-\nabla_{\vec{m}}w(\vec{m})/\mu_0$. 
$\vec{h}_1$ and $\vec{h}_2$ are in general non-collinear, thus 
the dynamics with the additional STT term in Eq. \eqref{LLG} 
is quite different\cite{xrw2} from that without this term 
which describes a Stoner particle in a magnetic field only:
The particle energy can only decrease in a static magnetic 
field since the field cannot be an energy source\cite{xrw}. 
However, a polarized electric current can pump energy into 
a Stoner particle through the STT. Thus, STT allows even a 
dc current to be an energy source, and dynamics and physics 
of a Stoner particle under a STT\cite{exp,zhang,sun}
are much richer than that under a static magnetic field. 
According to Eq. \eqref{DMLLG}, the magnetization undergoes 
a precessional motion around field $h_1$ and a damping 
motion toward field $h_2$.

In terms of polar angle $\theta$ and azimuthal angle 
$\phi$ of $\vec{m}$ in the spherical coordinates, Eq. 
\eqref{DMLLG} becomes
\begin{eqnarray}\label{sphe} 
& &(1+\alpha^2)\dot{\theta}=a_I(\alpha s_{\phi}-s_{\theta})-
\alpha(\frac{\partial w}{\partial\theta} - h_\theta)+\nonumber
\\ & &\qquad\qquad\quad h_\phi-\frac{1}{\sin\theta}
\frac{\partial w}{\partial\phi}\equiv F_1,\nonumber\\
& &(1+\alpha^2)\sin\theta\dot{\phi}=-a_I(\alpha s_{\theta}+
s_{\phi}) -h_\theta + \\
& &\qquad\qquad\frac{\partial w}{\partial
\theta}+ \alpha(h_\phi- \ \frac{1}{\sin\theta}
\frac{\partial w}{\partial\phi}) \equiv F_2, \nonumber 
\end{eqnarray}
where $\dot f$ means derivative of $f$ with respect to time $t$. 
Here $s_r$, $s_\theta$, $s_\phi$ are the $r,\ \theta$, 
and $\phi$ components of $\hat{s}$, and 
\begin{equation}\label{current} 
G_1\equiv s_\phi^2+s_\theta^2+s_r^2-1=0.
\end{equation}
Since the radial component of the external magnetic field 
does not appear in the dynamics equation of the magnetization. 
$h_r$ shall not affect the magnetization dynamics, and we 
shall always assume $h_r=0$ for an optimal field pulse. 
However, for STT induced reversal, $s_r$ will affect the 
magnetization dynamics through the STT coefficient $a_I$ which 
is a function of $P$ and $s_r$. In general, $F_i, \ (i=1,2)$ 
are functions of $\theta,\ \phi,\ h_\theta,\ h_\phi,\ s_\theta
,\ s_\phi$, and $s_r$. The switching problem is as follows: 
Before applying a field/current, magnetization $\vec m$ has two 
stable states, $\vec m_0$ (`north pole' $\theta=0$) and $-\vec 
m_0$ (`south pole' $\theta=\pi)$ along its easy axis (z-axis). 
Initially, the particle is in state $\vec m_0$, and the goal 
is to use a field pulse and a spin-polarized electric current 
pulse to switch the magnetization to $-\vec{m}_0$ quickly. 
Both the field direction and the spin polarization direction 
may vary with time. 

It will be beneficial to make a qualitative description 
of the relationships among field/current pulses, reversal 
trajectory, and reversal time. Given a field/current 
pulse, magnetization $\vec m$ will move following Eq. 
\eqref{sphe} if $\vec{m}_0$ is no longer a stable state 
under the field/current pulse. Eventually, the system 
will end up at one of its new stable states\cite{field}. 
If the pulse is proper, its evolution path $\phi(\theta)$ 
may pass through $-\vec{m}_0$. When that happens,  
the path is called a reversal trajectory, and the 
corresponding field/current pulses are reversal pulses. 
Given a reversal field/current pulse, there is a unique 
reversal trajectory which is the solution of Eq. \eqref
{DMLLG}, a nonlinear problem with no general analytic 
solution. However, knowing a reversal trajectory 
$\phi(\theta)$, which pass through $\theta=0, \pi$, 
there are infinite number of reversal pulses satisfying, 
according to Eq. \eqref{sphe}, 
\begin{equation}\label{path} 
\phi^\prime \sin\theta F_1-F_2\equiv G_2 =0, 
\end{equation}
where $\phi^\prime\equiv d\phi/d\theta$. 
All $h_i$ and $s_i$ ($i=r,\theta,\phi$) satisfying Eq. 
\eqref{path} form reversal pulses. 
Thus, there is a functional relationship between reversal 
pulse and reversal trajectory. What is more, 
given a reversal trajectory, magnetization ``velocities" 
$\dot\theta$ and $\dot\phi$ are linear in $\vec h$ and $I$. 
In other words, the reversal time shall be monotonic in $I$ 
and $h$, magnitude of $\vec h$. As $h$ and/or $I$ approach 
infinity, the reversal time goes to zero. Thus, the 
optimization problem (shortest reversal time) is meaningful 
only when one imposes additional conditions on $h$ and $I$. 
One sensible way is to fix $I$ and $h$, i.e., 
\begin{equation}\label{field}
G_3\equiv h_\phi^2+h_\theta^2-h^2=0.
\end{equation}

According to Eq. \eqref{sphe}, the reversal time $T$ 
along a reversal trajectory is given by 
\begin{equation}
T=\int_0^{\pi}\frac{d\theta}{\dot{\theta}}=
\int_0^{\pi}\frac{(1+\alpha^2)d\theta}{F_1}. 
\end{equation}
The optimization problem here is to find the optimal reversal 
trajectory for the fastest magnetization reversal under 
the constrains of Eqs. \eqref{current} and \eqref{field}.
Thus $T$ is minimal against the variation of the reversal 
trajectories $\phi(\theta)$ and field/current pulses $h_i$ 
and $s_i$ ($i=r,\theta,\phi$) that are linked together by 
Eq. \eqref{path}. This is just the Euler-Lagrange problem. 
Using Lagrange multipliers method, the optimal reversal 
trajectory and optimal reversal field/current pulses are 
given by 
\begin{eqnarray}\label{euler} 
& &\frac{d}{d\theta} (\frac{\partial F }{\partial \phi
^\prime})=\frac{\partial F }{\partial \phi}, \nonumber \\ 
& &\frac{d}{d\theta}(\frac{\partial F}{\partial h_i^\prime})=
\frac{\partial F}{\partial h_i},\qquad(i=\theta,\phi)\nonumber 
\\ & &\frac{d}{d\theta}(\frac{\partial F}{\partial s_i^\prime
})=\frac{\partial F}{\partial s_i},\qquad(i=r,\theta,\phi) 
\\ & &F=\frac{1+\alpha^2}{F_1}+\lambda_1G_1+\lambda_2 
G_2 + \lambda_3 G_3,\nonumber 
\\ & &G_1=0, \qquad G_2=0, \qquad G_3=0, \nonumber 
\end{eqnarray}
where $^\prime=\frac{d}{d\theta}$, thus, $f^\prime$ 
means derivative of $f$ with respect to $\theta$. 
$\lambda_{1,2,3}$ are the Lagrange multipliers that can 
be determined self-consistently from Eqs. \eqref{euler}.

Eqs. \eqref{euler} are the central result of the current work. 
For a given $I$, $P$, $h$, and particle ($w(\vec m)$ is 
known), the solution of the equations yield the optimal 
reversal trajectory $\phi(\theta)$ and optimal reversal 
field/current pulse $h_i$ and $s_i$ ($i=r,\theta, \phi$). 
Knowing the information, it is a straightforward task to find 
the theoretical limit of minimum switching field/current.  
Obviously, one can simultaneously optimize the 
field and the current in a controlled manipulation. 
From an application point of view, such a set up could be 
complicate and expensive, and one may be more interested in 
using one means only, i.e. either a field or a current, to 
manipulate a magnetization. Two cases may be interesting. 
Case A): Fix $I$ and $\hat s$, and vary $h_\theta$ and $h_\phi$. 
Reference 5 is a special example of the case with $I=0$. 
Case B): Fix the magnetic field $\vec h$ and the magnitude 
of current $I$, and vary the current polarization.
Reference 12 is a special example ($\vec h=0$) of the case. 

In order to demonstrate that Eq. \eqref{euler} is capable 
to find the optimal reversal trajectory for an arbitrary 
Stoner particle, let us consider case A), magnetic field 
induced magnetization reversal, $a_I=0$. After some tedious  
and straightforward calculation, the Euler equations become 
\begin{eqnarray}\label{Euler1} 
& &\frac{\partial^2 L }{\partial\phi^{\prime 2}}\phi^
{\prime\prime}+\frac{\partial^2L}{\partial\phi\partial 
\phi^\prime} \phi^\prime+\frac{\partial^2L}{\partial\theta 
\partial \phi^\prime}-\frac{\partial L}{\partial \phi}=0, 
\nonumber \\ & & f_1=\frac{\partial w}
{\partial\theta}-\frac{\alpha}{\sin\theta}\frac{\partial w}
{\partial\phi}; \quad f_2=-\alpha\frac
{\partial w}{\partial\theta} -\frac{1}{\sin\theta}
\frac{\partial w}{\partial\phi},\nonumber \\
& & L=\frac{\phi^\prime\sin\theta f_1+f_2}
{f_1^2+f_2^2-(1+\alpha^2)h^2} \pm 
\\ & & \frac{\sqrt{(1+\phi^{\prime 2}\sin^2\theta)
(1+\alpha^2)h^2 - (f_1-\phi^\prime\sin\theta f_2)^2}}
{f_1^2+f_2^2-(1+\alpha^2)h^2}, \nonumber 
\end{eqnarray}
where $f_{1,2}$ are functions of $\theta$ and $\phi$ 
determined by magnetic anisotropy energy $w(\vec m)$. 
$w=-km_z^2/2$ corresponds to References 5, and it can be 
shown that Eqs. \eqref{Euler1} is 
\begin{equation}
\phi^\prime=\frac{2\cos\theta}{\alpha(h/h_c-\sin 2\theta)}. 
\end{equation}
Its solution is 
\begin{eqnarray}
& & \phi=\phi_0+\frac{1}{\alpha}\sqrt{\frac{h_c}{h-h_c}}\arctan 
[\sqrt{\frac{2h_c}{h-h_c}}\sin(\theta-\frac{\pi}{4})]+ \nonumber
\\  & &\frac{\sqrt{h_c}}{2\alpha\sqrt{h+h_c}}\ln 
\frac{\sqrt{(h/h_c+1)/2}+\cos(\theta-\frac{\pi}{4})}
{\sqrt{(h/h_c+1)/2}-\cos(\theta-\frac{\pi}{4})},\nonumber  
\end{eqnarray}
where $\phi_0$ is an arbitrary constant reflecting the $\phi-$
symmetry of the system.  
This is exactly the same optimal trajectory equation obtained 
in Reference 5 with the theoretical limit of switching field 
being $h_c=max\{\alpha k\cos\theta\sin\theta/\sqrt{1+
\alpha^2}\}=\alpha k/(2\sqrt{1+\alpha^2})$. 
All other results in Reference 5 can be re-derived from this 
optimal reversal trajectory. 

For a biaxial Stoner particle of $w(\theta,\phi)=-(k_1/2)\cos
^2\theta + (k_2/2)\sin^2\theta \cos^2\phi$, Eq.\eqref{Euler1} 
may not be easy to solve analytically, but numerical 
solutions are easy to find. The solutions for $\alpha=0.1$, 
$k_1=1$, and $h=1 (\gg h_c$, which is about 0.05 for $k_2=0$) 
are presented in Fig. 1 for $k_2=0, 1, 5, 10$, respectively. 
Fig. 1a shows the optimal trajectories. It is interesting to 
note that the optimal trajectories for $k_2=0$ has rotational 
symmetry around z-axis (an infinite number of solutions). 
The one shown in Fig. 1a is just a particular solution with 
$\phi(t=0)=0$. For $k_2\neq 0$, the rotational symmetry is 
broken, but there are still four equivalent trajectories 
due to the biaxial symmetry. The ones whose inital $\phi$'s
are in the range of $[0,\pi/2]$ are used in the figure. 
Fig. 1b and 1c are the time evolution of $\theta$ and $\phi$ 
along the trajectories. It is surprising to note that 
switching time is shorter and shorter as 
$k_2$ increases, and the potential landscape is less smooth. 
Fig. 1d and 1e are the corresponding optimal reversal pulses. 
\begin{figure}[htbp]
\begin{center}\scalebox{0.80}[0.80]
{\includegraphics[angle=0]{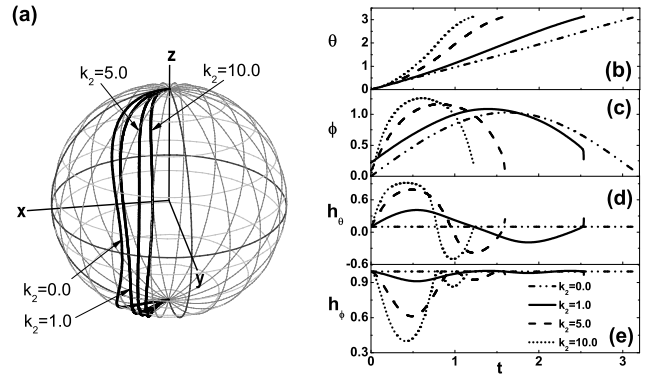}}
\end{center}
\vskip -0.20in
\caption{Numerical solutions of biaxial model with $\alpha
=0.1$, $k_1=1$, $h=1$, and various $k_2=0, 1, 5, 10$. 
a) The optimal reversal trajectories. b) $\theta$ vs. $t$.  
c) $\phi$ vs. t. d) $h_\theta$ vs. t. e) $h_\phi$ vs. t. }
\end{figure}
 
Similarly, one can obtain all the results in Reference 12 
from Eqs. \eqref{euler}. Only a spin-polarized current is 
used to manipulate the magnetization reversal of a uniaxial 
Stoner particle there, and the system specifities are $\vec
h=0$, $I=const.,$ $P=const.$, and $w=w(\cos\theta)$ with 
$-\partial w/\partial (\cos \theta)\equiv f(\cos\theta)$. 
After some algebras, the optimal trajectory equation is 
\begin{equation}
\phi^\prime= =\frac{f(\cos\theta)} 
{\alpha Q [I/I_c-f(\cos\theta)\sin\theta/Q]},
\end{equation}
where $I_c=\frac {\mu_0 e M^2 V}{\hbar G(P)}\frac{\alpha}
{\sqrt{1+\alpha^2}}Q$ is the theoretical limit of the 
critical switching current, $Q\equiv max\{f(\cos\theta)\sin
\theta\}$ for $\theta\in [0,\pi]$ depends on the magnetic 
anisotropy energy landscape, and $G(P)=g(P, s_r^*)\sqrt{1-
s_r^{*2}}$ comes from the $g$-function in Slonczewski STT.
The corresponding optimal polarization pulse is given by 
\begin{eqnarray}\label{pulse} 
& &s_r^{*2}=\frac{(1+P)^3}{16P^{3/2}-3(1-P)^3}, \nonumber \\
& &s_\phi^{*2}=-\alpha s_\theta^{*2}= 
\frac{\alpha}{\sqrt{1+\alpha^2}}\sqrt{1-s_r^{*2}}.
\end{eqnarray}
The trajectory equation has exactly the same form as that 
of magnetic-field case, reflecting the fact that both cases 
are described by the same dynamical equation \eqref{DMLLG}. 
These are exactly what were found in Reference 12 from a 
special observation. 

Our derivation of the Euler equation is based on the 
following observations. 1) Given a field/current pulse 
on a macro-spin of a magnetic nano-structure, the 
trajectory of its magnetization is uniquely determined 
by the corresponding LLG equation \eqref{sphe}. 
2) Knowing a reversal trajectory $\phi(\theta)$ 
(not $\theta(t)$ and $\phi(t)$ ), there are an infinite 
number of possible pulses, satisfying condition \eqref{path}, 
that can reverse the magnetization along the trajectory. 
3) The issue of the optimal magnetization reversal 
trajectory is an Euler-Lagrange problem with constrains. 

The Euler equation is more useful than the LLG equation 
in designing an optimal field/current pulse, although 
the Euler equation is derived from the later one. 
With only a LLG equation, one can only determine one 
magnetization trajectory for one specific field/current pulse. 
Thus, one does not even know whether the pulse can reverse 
a magnetization! Before obtaining all solutions of all possible 
pulses, one would not be able to tell which one is the best. 
With the Euler equation, one can obtain the optimal reversal 
trajectory and reversal pulses directly for a given particle. 
It should be pointed out that the current work address the 
same issues as those in References 5 and 12, but it provides 
a general and unified theory which can deal with combined 
effects of a field and a current for an arbitrary Stoner 
particle. 

In conclusion, a unified theory for the optimal magnetization 
reversal trajectories and reversal field and/or current 
pulses of an arbitrary Stoner particle is presented. 
The theory provides the Euler equation of the optimal 
reversal trajectory along which the magnetization reversal 
is the fastest is derived. Our early results on the critical 
switching field/current for a unixial Stoner particle are 
reproduced with the unified theory. The optimal magnetization 
reversal of a biaxial Stoner particle is also solved. 

{\it{Acknowledgments}--}This work is supported by UGC, 
Hong Kong, through RGC CERG grants (\#603106 and \#603007).

\end{document}